\documentclass[12pt,notitlepage]{utarticle-dan}

\usepackage{amsmath,amsthm,amsfonts,amscd,amssymb} 
\usepackage{eucal}
\usepackage{braket}
\usepackage{slashed}
\usepackage[hidelinks]{hyperref}
\usepackage{graphicx}
\usepackage{xcolor}
\usepackage{enumitem}
\usepackage{cite}
\usepackage{chapterbib}

\usepackage{authblk}
\usepackage{tikz} 
\usetikzlibrary{shapes,arrows,positioning,automata,backgrounds,calc,er,patterns,decorations.markings,decorations.pathreplacing,snakes,decorations.pathmorphing}
\usepackage{pgfplots}
\usetikzlibrary{quantikz2}

\newcommand{\be}{\begin{equation}}
\newcommand{\ee}{\end{equation}} 
\newcommand{\mb}{\mathbf}

\newlist{primenumerate}{enumerate}{1}
\setlist[primenumerate,1]{label={\arabic*$'$.}}

\numberwithin{equation}{section}

\tikzset{
photon/.style={decorate, decoration={snake}},
gluon/.style={decorate, draw=black,
    decoration={coil,amplitude=4pt, segment length=5pt}}
 }

\title{Comments on graviton detection}

    \author{Daniel Carney \\
	  Physics Division, Lawrence Berkeley National Laboratory \\
	  Berkeley, CA \\
	  {~}\\
	  \email{carney@lbl.gov}\\
       }

\Abstract{It is possible to make a detector which clicks after absorbing a single graviton. Similarly, it is possible to make a gravitational wave detector which can see the quantum noise induced by certain highly squeezed states of the graviton. However, contrary to some recent arguments in the literature, observation of either or both of these signals would not constitute proof or even evidence that the gravitational field is quantized. This is a simple technical statement: a classical gravitational wave can produce the same output data in a detector. Here we explain this result, presented earlier in \cite{Carney:2023nzz}, which is a straightforward extension of an ancient argument in quantum optics. \\ \\ Dedicated to Gordon Semenoff on the occasion of his 70th birthday, and based on a talk at his festschrift held at the Universit\'e de Montr\'eal, July 2023.}

\begin{document}

\maketitle

\newpage

\section{Introduction}

By simple analogy with the other known fields of nature, it seems highly likely that the gravitational field is quantized, but experimental evidence for this idea is currently lacking. Theoretically, it is straightforward to perturbatively quantize general relativity around a background and treat the resulting graviton fluctuations as an effective quantum field theory \cite{Donoghue:2022eay}. This model makes very definite predictions, and these predictions can be measured in real experiments, as discussed below. However, the right question is whether we can form a reasonable alternative to the graviton picture and perform an experiment that can distinguish the two cases, i.e., do a measurement that would provide us with more Bayesian evidence that the gravitational field is quantized. As we will see, this is a substantially harder goal than simply confirming some prediction of graviton physics.

What kind of experiment could one do to test the quantization of gravity, i.e., the existence of the graviton? A seemingly obvious method would be to just ``detect a graviton''. It may sound like it is impossibly difficult to detect single gravitons, but in the modern understanding of quantum measurement, it is actually pretty straightforward. Detecting a graviton would mean constructing a detector which clicks when it absorbs a graviton, much as a photodiode clicks when it absorbs a photon. It is easy to think of ways to do this. Examples include absorption through a quadrupole transition in an atom \cite{Boughn:2006st}, using single-photon counters after converting the gravitons into photons through the Gertsenshtein effect \cite{Dyson:2013hbl,Palessandro:2024ria}, or using single-phonon counters after converting the gravitons into phonons in a Weber bar \cite{Tobar:2023ksi,Shenderov:2024rup}.

As remarkable as it would be to see such an event, we now have to ask the right question: would such a detection provide evidence for the existence of the graviton? The answer, unfortunately, is not really \cite{Carney:2023nzz}. Consider the following pair of models for a detector coupled to an incident gravitational signal:
\begin{align}
\label{H-q}
\hat{H}_{\rm q} & = \hat{H}_{\rm det} - \int d^3\mb{x} \hat{h}_{\mu\nu} \hat{T}^{\mu\nu}_{\rm det} \\
\label{H-cl}
\hat{H}_{\rm cl} & = \hat{H}_{\rm det} - \int d^3\mb{x} h_{\mu\nu} \hat{T}^{\mu\nu}_{\rm det}.
\end{align}
In both models, the detector itself is treated as a quantum system. The difference is whether or not the gravitational field perturbation $h_{\mu\nu}$ is treated as a quantum field (denoted by a hat) or as a classical field (without a hat). In general, we might imagine that the classical field $h_{\mu\nu}$ is drawn from a classical random distribution $P_{\rm cl}(h_{\mu\nu})$. We are not specifying the internal dynamics of $h$, just treating it as some external signal incident on the detector. Here is the claim: any realistic ``graviton'' signal picked up by such a detector can equally well be explained by \eqref{H-q} or \eqref{H-cl}.\footnote{Some attempts to make self-consistent dynamical models in the classical case include \cite{Kafri:2014zsa,Tilloy:2018tjp,Oppenheim:2023izn}. Taken literally, Eqs. \eqref{H-q} and \eqref{H-cl} suggest violations of energy conservation, since there is no loss of energy in the field after detection, as emphasized in \cite{Tobar:2023ksi,Shenderov:2024rup}. However, this is irrelevant to the point of this paper, because the state of the field after it interacts with the detector is not measured. Thus this violation of energy conservation cannot be seen in the experiments described here.}  

In what follows, we will make this statement precise.  We begin by discussing the analogous statement regarding the electromagnetic field and photon detection, which is an almost identical argument \cite{Glauber:1963tx,sudarshan1963equivalence,mandel1995optical}. First we cover the case of photon detection and discuss what kinds of events can and cannot be explained by a classical model of electromagnetic radiation. We then extend these arguments to the gravitational case. In both the electromagnetic and gravitational cases, we emphasize that similar arguments apply to both ``counting'' detectors (like photodiodes or atomic absorbers) and to ``linear'' or ``amplitude'' detectors (like homodyne detectors and LIGO, which measures strain $h_{\mu\nu}$ directly). The latter case will clarify why measuring things like ``quantum noise'' in squeezed states with a detector like LIGO \cite{Guerreiro:2019vbq,Parikh:2020fhy} can similarly be explained by a classical radiation model except in very extreme, and physically unrealistic, scenarios. We emphasize this with an explicit calculation in LIGO in Sec. \ref{ligo}, which is the only new material appearing here that was not essentially in Ref. \cite{Carney:2023nzz}.

\section{Photoelectric effect}

Consider the electromagnetic analogues of the two models \eqref{H-q} or \eqref{H-cl}:
\begin{align}
\label{H-q-EM}
\hat{H}_{\rm q} & = \hat{H}_{\rm det} - \int d^3\mb{x} \hat{A}_{\mu} \hat{J}^{\mu}_{\rm det} \\
\label{H-cl-EM}
\hat{H}_{\rm cl} & = \hat{H}_{\rm det} - \int d^3\mb{x} A_{\mu} \hat{J}^{\mu}_{\rm det}.
\end{align}
Again the detector is a quantum, dynamic system; the electromagnetic field is being treated as non-dynamical, external perturbation; and the difference between the models is whether the electromagnetic field is quantized $\hat{A}$ or a classical variable $A$ drawn in general from some probability distribution $P_{\rm cl}(A)$. The question is what kind of state $\ket{\psi}$ of the quantized electromagnetic field and detector observable could be used to distinguish between the two models \eqref{H-q-EM} and \eqref{H-cl-EM}.

Let's begin with a famous \emph{non}-example \cite{Glauber:1963tx,sudarshan1963equivalence,mandel1995optical}: detection of individual clicks in a photodiode based on the photoelectric effect. We model the photodiode $\hat{H}_{\rm det}$ as a bunch of electrons, each with a spectrum
\be
\hat{H}_{\rm det} = -\Delta \ket{g} \bra{g} + \sum_{E=0}^{\infty} E \ket{E} \bra{E}.
\ee
Here $\Delta > 0$ is the energy gap between the ground state, in which the electron is bound to the atomic lattice, and the conduction band states $\ket{E}$ representing unbound electrons (often then referred to as ``photoelectrons''). We are treating this as a discrete spectrum with no degeneracy for simplicity. Properly we should have a sum over $N \gg 1$ individual electrons in $\hat{H}_{\rm det}$ but for notational simplicity we will suppress this. 

\begin{figure}[t]

\centering

\begin{tikzpicture}[scale=0.5]

\draw (-3,1) rectangle (3,-1);

\draw [fill=black] (-2,.5) circle (.1);
\draw [fill=black] (-1.1,.2) circle (.1);
\draw [fill=black] (-.2,-.4) circle (.1);
\draw [fill=black] (1.2,0) circle (.1);
\node at (-2,-.3) {$\ket{g}$};

\draw [->] (1.5,.3) -- (3,2);
\node at (4.3,2) {$\ket{E}$};

\draw [photon,thick,->] (-4,3.5) -- (-2,1.5);
\draw [photon,thick,->] (-3,3.5) -- (-1,1.5);

\node at (-2,4) {$\omega$};

\begin{scope}[shift={(15,0)}]

\draw[dotted,photon,thick] (1.3,0) -- (7,0);
\node at (4,-1) {$a$};

\draw [thick,->] (-2,-3) -- (-2,3);
\draw [thick,->] (0,-3) -- (0,3);
\draw [thick,->] (2,-3) -- (2,3);

\node at (3,3) {$\mathbf{B}_0$};

\draw [black] (7,0.5) -- (7,-0.5) arc(-90:90:.5) -- cycle;

\draw[photon] (-7,0) -- (0.7,0);
\node at (1,0) {$\otimes$};
\node at (-5,-1) {$b$};

\draw[->] (-6,1) -- (-4,1);
\node at (-5,2) {$\mb{k}$};
\end{scope}
\end{tikzpicture}
\caption{Left: a single-photon detector, based on the photoelectric effect. Incident electromagnetic radiation excites bound electrons $\ket{g} \to \ket{E}$, which are then sent through an amplifier and detected. Right: a single-graviton detector, based on the Gertsenshtein effect. Incident gravitational radiation is converted to X-ray frequency light through the interaction $S = \int d^4x T_{\mu\nu}^{EM} h^{\mu\nu}$ in the presence of a large external field $\mathbf{B}_0$, and the subsequent X-ray photons are then detected.}
\label{fig:diodes}
\end{figure}
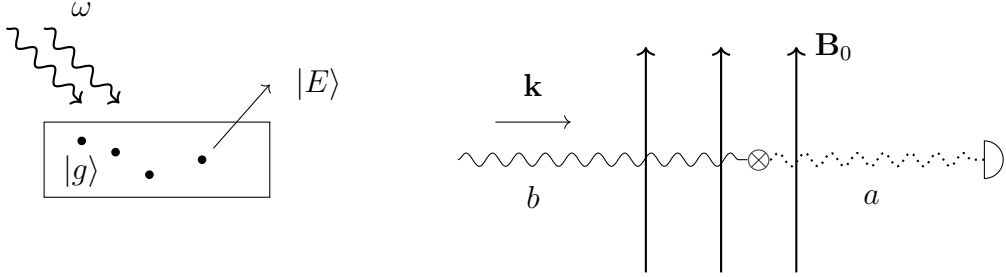

The detector ``clicks'' when an electron transitions from $\ket{g} \to \ket{E}$ for any $\ket{E}$. In the quantized radiation model \eqref{H-q-EM}, this is explained by the electron absorbing a photon. Historically, this explanation and the observation of individual photoelectron excitations was viewed as evidence for the quantization of the electromagnetic field \cite{https://doi.org/10.1002/andp.19053220607}. However, these clicks can equally well be explained by both \eqref{H-q-EM} and \eqref{H-cl-EM}. They both lead to precisely the same formula for the rate of photoelectron excitation, as we now show explicitly.

We can start with the classical case. Consider for simplicity a monochromatic incident field $\mb{A}(\mb{x},t) = \mb{N}_0 \left[ b e^{i k \cdot x} + {\rm h.c.}\right]$, where $\mb{N}_0$ is some fixed vector, $b$ is the (complex) Fourier amplitude for the mode, $\mb{k}$ is the wavevector and $\omega = |\mb{k}|$ is the frequency, and we are working in $A^0 = 0$ gauge. The electron has $\hat{\mb{J}}(\mb{x}) = (e \hat{\mb{p}}/m) \delta^3(\mb{x}-\hat{\mb{x}})$, where $\hat{\mb{x}}$ is the position operator of the electron. Assuming the electron is localized near the origin to better precision than the wavelength of the incident field $\lambda = 1/\omega$, the $\mb{J} \cdot \mb{A}$ coupling takes the simple form
\be
\hat{V}_{\rm cl}(t) = - \frac{e}{m} \hat{\mb{p}} \cdot \mb{N}_0 \left[ b e^{-i \omega t} + {\rm h.c.} \right].
\ee
The amplitude for an electron to excite into a given conduction state $\ket{E}$ in some small time interval $\delta t$ is given by the usual formula from first-order perturbation theory
\begin{align}
\begin{split}
\label{M-q}
\mathcal{M}(g \to E) & = -i \int_{-\delta t/2}^{\delta t/2} dt \ \braket{E |  V_{{\rm cl,}I}(t) | g} \\
& = i \frac{e}{m} \int_{-\delta t/2}^{\delta t/2} dt \ \mb{N}_0 \cdot \braket{E| \hat{\mb{p}} | g} \left[  b e^{-i (E + \Delta - \omega) t} +   b^* e^{-i (E + \Delta + \omega) t} \right] \\
& \approx i \frac{2e}{m} \mb{N}_0 \cdot \braket{E| \hat{\mb{p}} | g} b \frac{ \sin \left[ \left( E + \Delta - \omega\right) \delta t/2 \right]}{E + \Delta - \omega}.
\end{split}
\end{align}
In the first line, the subscript $I$ means interaction picture. To get to the last line, we assumed that $\delta t \gtrsim 1/\omega$, in which case the exponential in the first term integrates to the sinc function, while the exponential in the second term averages to zero. This is an excellent approximation in an optical photodiode for example, where $\omega \sim 0.1~{\rm eV} \sim 100~{\rm THz}$ and even a fast photodiode would only bin the events at something like $\delta t \sim 1~{\rm ns}$ rates. The probability that the detector clicks in this time window is given by squaring this amplitude and summing over all possible conduction band states, which gives
\be
\label{dP-cl}
\frac{d\mathcal{P}(g \to {\rm any}~E)}{dt} = \eta I_b \Theta(\omega - \Delta),
\ee
where we took the limit $\omega \delta t \gg 1$ to write the sinc$^2$ as $\delta(E + \Delta - \omega) \delta t$, and defined the incident intensity $I_b$ (given the Fourier amplitude $b$) and detector efficiency $\eta$ as
\begin{align}
\begin{split}
I_b & = |b|^2 \\
\eta & = \frac{e^2}{m^2} \rho(\omega-\Delta) | \mb{N}_0 \cdot \braket{\omega-\Delta | \hat{\mb{p}} | g}|^2.
\end{split}
\end{align} 
We took the continuum limit of the conduction states $\ket{E}$, assigning them a density of states $\rho(E)$. The dimensionless quantity $\eta$ gives the fraction of the power of an incident beam polarized along $\mb{N}_0$ which is absorbed by the detector. Modern photodiodes can achieve $\eta \gtrsim 0.9$, meaning they absorb $\gtrsim 90\%$ of photons in a beam aimed at them.

The quantum calculation is very similar. Consider the same electromagnetic mode $(\mb{k},\omega)$, which we now expand into creation and annihilation operators
\be
\hat{\mb{A}}(\mb{x},t) = \mb{N}_0 e^{i k \cdot x} \hat{b} + {\rm h.c.},
\ee
where $\hat{b}$ is the annihilation operator for the mode and again $\mb{N}_0$ is a fixed overall vector. Now consider an incident coherent state $\hat{b} \ket{\beta} = \beta \ket{\beta}$ of this mode; we will generalize to an arbitrary state shortly. The amplitude of interest now has to allow for the field state to change. Let us assume that the state of the electromagnetic field is not measured after the interaction with the detector. Then the probability that a photoelectron is excited into a specific state $\ket{E}$ comes from averaging over the final states of the field,
\begin{align}
\begin{split}
\mathcal{P}(g \to E) & = \int d\alpha \left| \mathcal{M}(g, \beta \to E,\alpha) \right|^2 \\
& \approx \frac{e^2}{m^2} \left| \mb{N}_0 \cdot \braket{E| \hat{\mb{p}} | g} \right|^2 |\beta|^2 \frac{ \sin^2 \left[ \left( E + \Delta - \omega\right) \delta t/2 \right]}{(E + \Delta - \omega)^2}.
\end{split}
\end{align}
Here $\ket{\alpha}$ is another coherent state. The calculation of $\mathcal{M}$ is identical to that in \eqref{M-q}; we used the same arguments to drop the $e^{i(E + \Delta + \omega)t}$ term, as well as the completeness relation of the coherent states $\int d\alpha \ket{\alpha} \bra{\alpha} = \pi$. Finally, we can integrate over the possible conduction states to get the total rate of probability for a click,
\be
\label{dP-q}
\frac{d\mathcal{P}(g \to {\rm any}~E)}{dt} = \eta I_{\beta} \Theta(\omega-\Delta),
\ee
where now $I_{\beta} = |\beta|^2$ is the intensity of the light in the state $\ket{\beta}$ and the efficiency $\eta$ is the same as before. 

The conclusion is straightforward. The prediction for the rate of photoelectron clicks in the detector is the same in either the model with quantized radiation or simple classical radiation, once one makes the identification $b \to \beta$. Notice all the features that are commonly taught as being due to the quantum nature of light persist: the click rate is proportional to the intensity of the light $I$, the clicks only occur when ``energy is conserved'' $\omega \geq \Delta$ (which in the classical case is a simple resonance condition between the light and the states above the electron gap), and the clicks themselves are of course discrete. 

It should be emphasized that we are not saying that a purely classical model of the universe can reproduce the photoelectric effect. Rather, we are saying that a classical model of the \emph{radiation field}, coupled to a quantum model of the detector, can reproduce the photoelectric effect. In the context of electromagnetism this is purely academic since we have many other tests of the quantization of the radiation field. However, this is \emph{precisely} the situation in a gravitational wave detector at the present time: we know that the detectors are made of normal quantum matter, but we are looking for evidence that the gravitational radiation field itself is quantized.

Here we presented these rates with either a fixed-amplitude classical field \eqref{dP-cl} or quantum field \eqref{dP-q}. This can be easily generalized to an arbitrary state in either case. For the classical case, we want to allow for a randomly drawn field amplitude, say from a distribution $P_{\rm cl}(b)$. In the quantum case, we can instead use the Glauber-Sudarshan representation: for any state $\rho$ of a harmonic oscillator, one can write
\be
\label{glauber}
\rho = \int d\beta P(\beta) \ket{\beta} \bra{\beta},
\ee
where the weight function $P(\beta)$ is real, formally defined as the integral \cite{Glauber:1963tx,mehta1967diagonal}
\be
\label{mehta}
P(\beta) = \frac{e^{|\beta|^2}}{\pi^2} \int d\alpha \braket{-\alpha | \rho | \alpha} e^{|\alpha|^2 - \beta^* \alpha + \beta \alpha^*} 
\ee
and, when it exists, satisfies $\int d\beta P(\beta) = 1$. It is easy to track through the calculations of \eqref{dP-cl} and \eqref{dP-q} to see that in an arbitrary state, either classical or quantum, the click rate is 
\be
\label{dP-both}
\frac{d\mathcal{P}}{dt} = \eta \braket{ I } \Theta(\omega-\Delta),
\ee
with the only difference being whether the bracket means a classical expectation $\braket{I} = \int db P_{\rm cl}(b) I_b$ or a quantum expectation $\braket{I} = \int d\beta P(\beta) I_{\beta}$. 

\section{Non-classicality tests by counting photons}

What Eq. \eqref{dP-both} shows is that the expected rate $d\mathcal{P}/dt$ of discrete photoelectron detector clicks can always be explained equally well with a classical or quantum model of the electromagnetic radiation. In this sense, observation of the photoelectron effect does not demonstrate that the photon exists. The extension to gravitational radiation is immediate. The question, then, is what kind of observation \emph{would} distinguish the classical and quantum models? 

Generally speaking, what has been understood in quantum optics is that one can make sharp tests based on correlation statistics beyond averages, and given certain quantum states of the radiation field. This is similar in spirit to Bell inequality violations: it is possible to find correlation functions which are bound by some quantity in any classical model, but for which certain quantum states violate the bound. In the next subsections, we give explicit examples based on click counting statistics (e.g., variance in the click rate) and noise statistics (e.g., variance in the observed electric field strength). 

First, we give a characterization of the types of ``non-classical'' quantum states of radiation we will be interested in. In a given quantum state $\rho$, the Glauber function $P(\beta)$ may or may not be positive for all $\beta$. If $P(\beta) \geq$ everywhere, then $P(\beta)$ is a probability distribution, so the state \eqref{glauber} can be interpreted as a classical ensemble of quantum coherent states or classical states of definite Fourier amplitude. However, if $P(\beta) < 0$ for some region of $\beta \in \mathbb{C}$, then $P(\beta)$ is not a probability distribution, and it turns out that in this case one can find observables that can only be explained by a quantum mechanical model of the radiation field. Thus people somtimes refer to states whose Glauber-Sudarshan representation is somewhere negative as ``non-classical''. This terminology is obviously a bit loaded. A more conservative convention would be to say that these states have Glauber negativity. Similar terminology is often used in the Wigner function representation.

Before moving on to gravity, let's discuss a concrete example of a ``photon counting'' measurement that can distinguish between \eqref{H-q-EM} and \eqref{H-cl-EM}. Suppose that we shine light on our photodetector for some time $T$ and measure the total number of photoelectron clicks, $N$. If we do this many times, then from the measured values $N_1, N_2, \dots, N_{n_{\rm runs}}$ we can calculate the average $\overline{N} = \sum_i N_i/N_{n_{\rm runs}}$ and variance $\Delta N^2 = \sum_i (N_i - \overline{N})^2/N_{n_{\rm runs}}$. Here is the statement: an arbitrary classical state with classical random distribution $P_{\rm cl}(b)$ will always produce Poisson or super-Poisson statistics, $\Delta N^2 \geq \overline{N}$. However, certain ``non-classical'' states, for example squeezed states, can produce sub-Poisson statistics where $\Delta N^2 < \overline{N}$. Thus, observation of sub-Poisson click rates would necessarily imply that the electromagnetic field has to be quantized.

The proof is by direct calculation. First consider the classical case, where the Fourier amplitude in general is drawn from some distribution $P_{\rm cl}(b)$. In each run of the experiment, we get some realization of this distribution, and it produces a Poisson-distributed set of clicks, with expected value $\overline{N}_b = \eta I_b T$ and variance $\Delta N_b^2 = \overline{N}_b$. Here the statistics come from the quantum nature of the \emph{detector electrons} being measured. Now, performing many runs, each draws from $P_{\rm cl}(b)$, which leads to even more randomness since we are averaging over multiple Poisson distributions. One finds \cite{Carney:2023nzz,mandel1995optical}
\be
\Delta N^2 = \overline{N} + \eta^2 T^2 \int db P_{\rm cl}(b) \left[ I_b - \braket{I} \right]^2.
\ee
Here, the first term $\overline{N} = \int da P_{\rm cl}(b) \eta I_b T$ is the average number expected over all the runs. The second term is just the variance of the incident electromagnetic field intensity $\braket{\Delta I^2}$. This is positive, so this equation shows that the observed click rates are always either Poisson or super-Poisson $\Delta N^2 \geq \overline{N}$. The noisier the input field (larger $\Delta I^2)$, the noisier the data.

Now consider the quantum model of radiation. A directly analogous calculation can be performed in an arbitrary state with Glauber-Sudarshan representation $P(\beta)$. In a given coherent state, as discussed above, the click statistics will be Poisson distributed with $\overline{N}_{\beta} = \eta I_{\beta}T$ and $\Delta N_{\beta}^2 = \overline{N}_{\beta}$. Now, with many runs, one might try to repeat the same argument as the classical case, averaging over $P(\beta)$. But in general $P(\beta)$ is not a true probability distribution, so we have to be slightly more careful, and in particular operator ordering changes the answer slightly. One finds \cite{Carney:2023nzz,mandel1995optical}
\be
\label{N-q}
\Delta N^2 = \overline{N} + \eta^2 T^2 \int d\beta P(\beta) \left[ I_{\beta} - \braket{I} \right]^2.
\ee
The first term is just $\overline{N} = \int d\beta P(\beta) I_\beta T$ as in the classical case. However, the intensity $I = b^\dag b$ as an operator, so the quantum expectation $\braket{I^2} = \braket{b^\dag b b^\dag b} \neq \int d\beta P(\beta) I_{\beta}^2 = \int d\beta P(\beta) |\beta|^4$, because one has a non-trivial contribution from the $[b,b^\dag]=1$ commutator. Less prosaically, the second term in Eq. \eqref{N-q} is not the variance of an operator, and in particular, if $P(\beta) < 0$ for some $\beta$, it is possible that this integral is \emph{negative}. This is the case for example in for phase-squeezed states of the radiation field \cite{loudon1987squeezed}. Thus quantum radiation fields allow for sub-Poisson statistics $\Delta N^2 < \overline{N}$. 

Sub-Poisson statistics of the electromagnetic field were first experimentally observed in the 1980's \cite{short1983observation}. They now form a routine method of reducing shot noise in practical experiments \cite{Jia:2024epj}. This observation provides very definitive evidence that the electromagnetic field is quantized, since this observation cannot be reproduced by any classical distribution of radiation fields.

\section{Non-classicality tests with gravitational radiation}
\label{sec-gravity}

Extending the arguments presented above to the case of gravitational wave or ``graviton'' detectors is mostly straightforward. The question is what kind of detection could distinguish a model where the gravitaitonal radiation field is classical or quantized, in the sense of the models in Eqs. \eqref{H-q} and \eqref{H-cl}. The only subtlety is that most gravitational wave detectors currently do not measure the gravitational wave intensity---i.e., they do not actually count gravitons, but rather measure the strain waveform as a function of time. I will discuss the difference below.

First, however, we can quickly cover the case of an actual graviton counting detector, such as those suggested in \cite{Boughn:2006st,Dyson:2013hbl,Tobar:2023ksi}. Here we have a photodiode-like device which ``clicks'' on absorption of a single graviton. The question is whether a classical gravitational model can reproduce the same clicks. The answer is the same as in the discussion of the photoelectric effect above. The simple counting of such clicks proves nothing, i.e., cannot distinguish between the models \eqref{H-q} and \eqref{H-cl}. On the other hand, one might try to do something like observe sub-Poisson graviton counting statistics, as in Eq. \eqref{N-q}. This would first of all require a source of squeezed gravitational radiation, which would be exotic. Worse, though, it would be unobservable, for a simple reason. Notice in Eq. \eqref{N-q}, the deviation from Poisson statistics is proportional to the efficiency:
\be
\Delta N^2 - \overline{N} \propto \eta^2 \sim \left( \frac{P_{\rm absorbed}}{P_{\rm incident}} \right)^2.
\ee
In a state-of-the-art photodiode, $\eta \sim O(1)$ is possible. But in a typical gravitational wave detector, $\eta \ll 1$ is astronomically small. In the CAST experiment, for example, $\eta \sim 10^{-33}$ \cite{Carney:2023nzz}. What this means is that any ``graviton detector'' is going to see a Poisson distribution of clicks, and these can equally well be explained by classical gravitational radiation. The intuition is that gravity is weakly coupled to the detector, so the measurement is very weak, and we are trying to determine the underlying sub-Poisson statistics of the signal using an extremely minute sample set, which is impossible. The statistics are swamped instead by the detector's own noise.

What about strain detectors like LIGO? Here the observable is not the incident intensity $I \sim b^\dag b$ but rather the incident strain field $h \sim b + b^\dag$. These are ``linear'' detectors rather than ``square law'' detectors of the gravitational field. Non-classicality tests in linear observables are also available. The rough idea is that the two variables $h$ and its conjugate $\pi_h \sim -i(b - b^\dag)$ come with uncertainties $\Delta h = \Delta \pi_h = 1/2$ in the vacuum or a coherent state. By similar logic as above, a detector registering this level of noise could just as well be explained with an incoming classical wave. However, certain quantum states like squeezed states can achieve sub-vacuum levels of noise, say $\Delta h < 1/2$ (while $\Delta \pi_h > 1/2$ by Heisenberg uncertainty). It turns out that there are observables analogous to sub-Poisson statistics in such states that cannot be explained classically. However, much like sub-Poisson statistics, detecting these would require an impossibly efficient gravitational wave detector.

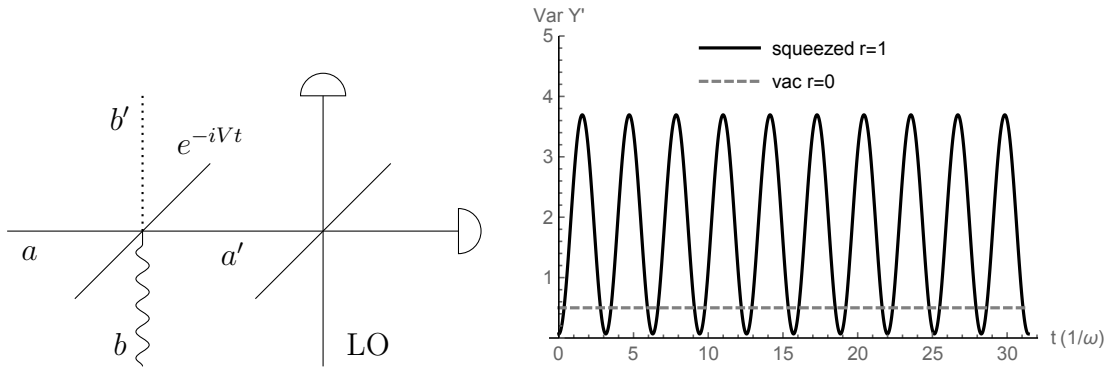
\begin{figure}[t]
\centering

\begin{tikzpicture}[scale=.6]

\draw (-3,0) -- (0,0);
\draw [photon] (0,-3) -- (0,0);
\draw [dotted,thick] (0,0) -- (0,3);
\draw (0,0) -- (4,0);

\node at (-2.5,-.5) {$a$};
\node at (-.5,-2.5) {$b$};
\node at (-.5,2.5) {$b'$};
\node at (2,-.5) {$a'$};

\draw (-1.5,-1.5) -- (1.5,1.5);

\node at (1.5,2) {$e^{-i V t}$};

\draw (4,-3) -- (4,0);
\node at (5,-2.5) {LO};
\draw (4,0) -- (7,0);
\draw (4,0) -- (4,3);

\draw (2.5,-1.5) -- (5.5,1.5);

\draw [black] (3.5,3) -- (4.5,3) arc(0:180:.5) -- cycle;

\draw [black] (7,0.5) -- (7,-0.5) arc(-90:90:.5) -- cycle;
\end{tikzpicture} ~ ~
\includegraphics[width=.5\linewidth]{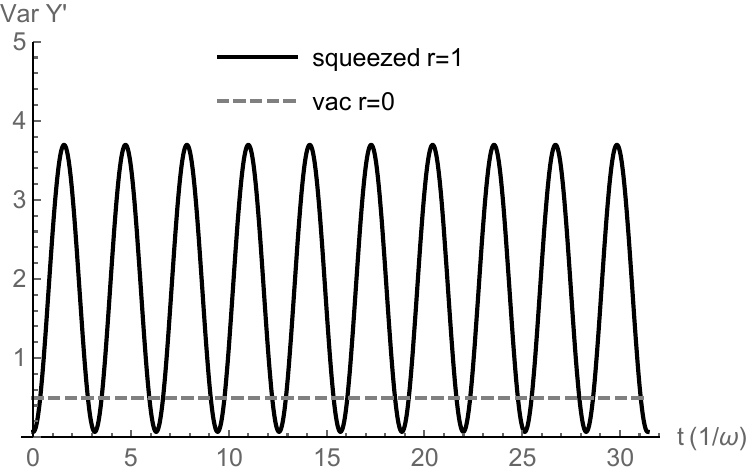}
\caption{Left: simple beamsplitter model of the interaction [first diagonal line, labeled $e^{-i V t}$, as in Eqs. \eqref{Vint} and \eqref{Vcl}], followed by a homodyne measurement to read out $Y' = -i (a' - a^{'\dag})/\sqrt{2}$. The quantum state of the signal mode after the interaction ($b'$) is lost to the bath. Right: Variance in the detector quadrature output $\Delta Y^{'2}(t)$ as a function of time, with an incoming squeezed signal in a mode of frequency $\omega$. We see that the noise oscillates between values above and below the vacuum noise value $1/2$. The large noise above the vacuum can be equally well explained with a \emph{classical} signal, as discussed in the text. While the squeezed state does produce a large amount of noise (compared to vacuum), only the small sub-vacuum components require a quantum-mechanical signal to explain them.}
\label{fig:time-dep}
\end{figure}

Let's give a simple, precise version of this idea as described in \cite{Carney:2023nzz}, and then connect it to more standard language involving noise power spectra of gravitational waves. See Fig. \ref{fig:time-dep}. Consider two modes, call them $a$ and $b$, which interact through the linear exchange interaction
\be
\label{Vint}
V = -i g (a b^\dag - a^\dag b).
\ee
Here $a$ will be the mode we ultimately detect, while $b$ is the mode with the signal of interest --- either the gravitational wave or some electromagnetic mode. This is the effective interaction generated by, for example, the coupling $V = \int d^3\mb{x} h_{\mu\nu} T^{\mu\nu}$ once linearized around a large mass detector \cite{Beckey:2023shi}. The interaction \eqref{Vint} generates a beamsplitting effect, in the sense that the Heisenberg equations $\dot{a} = i [V,a]$ and $\dot{b} = i [V,b]$ have the solutions
\be
\label{atbt}
a(t) = a \cos (gt) - b \sin(gt), \ \ b(t) = b \cos (gt) + a \sin(g t).
\ee
For $gt=\pi/4$, this is a 50/50 beamsplitter. For $gt \ll 1$, this is a mostly transmissive mirror. Suppose that we send in vacuum $\ket{0}$ to the detector mode $a$, and a squeezed state $S(r,\theta) \ket{0}$ in the signal mode $b$, where the squeezing operator is
\be
S(r,\theta) := \exp \left\{ \frac{1}{2} \left( r e^{-i \theta} b^2 - r e^{i \theta} b^{\dag 2} \right) \right\}.
\ee
Now, we are going to detect the mode $a'$ after the initial beamsplitter, say using a homodyne detector to measure $Y' = -i (a' - a^{'\dag})/\sqrt{2}$ (the ``phase quadrature'', i.e., the phase of the light exiting the beamsplitter). See Fig. \ref{fig:time-dep}. We want to know what the variance of the output will be, as a function of the signal squeezing $r$.

To calculate this, we can work out how the squeezing and beamsplitter operations transform the various modes. Let's take $\theta = \pi$, i.e., the signal is squeezed in the signal's phase quadrature $h = -i (b - b^\dag)/\sqrt{2}$, which will connect to our discussion of LIGO below. We have
\be
S^\dag b S = b \cosh r + b^\dag \sinh r, \ \ \ S^\dag b^\dag S = b^\dag \cosh r + b \sinh r.
\ee
Thus the detector mode, after mixing with this squeezed $b$ mode, has phase
\be
Y' = -i \frac{a - a^\dag}{\sqrt{2}} \cos (g t) - i \frac{b - b^\dag}{\sqrt{2}} e^{-r} \sin (g t),
\ee
where we used Eq. \eqref{atbt}. The variance of the measurements is then given by
\be
\Delta Y'^2 = \braket{ Y^2 } \cos^2(g t) + \braket{\pi_h^2} e^{-2 r} \sin^2(gt) = \frac{1}{2} \left[ \cos^2(g t) + e^{-2 r} \sin^2(gt) \right],
\ee
where the expectation values are taken in vacuum. For zero squeezing $r = 0$, this is just $\Delta Y'^2 = 1/2$, the vacuum value. This is the case where both $a,b$ are in vacuum (or coherent) states, mix at the beamsplitter, and therefore produce the same amount of output noise. For non-zero squeezing $r > 0$, however, the observed variance $\Delta Y'^2$ will oscillate in time between sub-vacuum $\Delta Y'^2 < 1/2$ and super-vacuum $\Delta Y'^2 > 1/2$ levels of noise. For comparison to the case of LIGO below, it is helpful to consider the $gt \ll 1$ limit, where the signal is very weakly coupled to the detector. In this case, 
\be
\label{deltaXpert}
\Delta Y'^2 = \frac{1}{2} \left[ 1 - (1 - e^{-2 r}) g^2 t^2 + \cdots \right] \leq \frac{1}{2},
\ee
with equality only for $r = 0$. What is happening is that the sub-vacuum signal (the squeezed noise in $b$) is getting transduced onto the detector mode $a$. Note that by unitarity, this means that extra noise is now pushed to the signal mode $b$, which is not measured.

What if the mode $b$ is treated purely classically? In this case, the interaction \eqref{Vint} becomes a simple unitary on the detector mode $a$, namely
\be
\label{Vcl}
V_{\rm cl} = -i g (a \beta^* - a^\dag \beta),
\ee
where $\beta$ is a c-number, possibly drawn from a classical distribution $P_{\rm cl}(\beta)$ as above. The detector mode after this interaction is
\be
Y' = -i\frac{a - a^\dag}{\sqrt{2}} - \frac{\beta + \beta^*}{\sqrt{2}} g t,
\ee
which is just the original operator $X$ operator shifted by a c-number. In particular, its variance is
\be
\Delta Y'^2 = \frac{1}{2} + \Delta h^2_{\rm cl}, \ \ \ \Delta h^2_{\rm cl} = \int d\beta P_{\rm cl}(\beta) \frac{(\beta + \beta^*)^2}{2} g^2 t^2.
\ee
We see that the classical signal can produce vacuum or super-vacuum output noise $\Delta Y'^2 \geq 1/2$, but not sub-vacuum noise $\Delta Y'^2 < 1/2$. The analogy with Poisson and sub-Poisson statistics should be clear.

As a final detail, we note that these calculations were performed as if the two modes $a,b$ have no time dependence other than the interaction, i.e., we are in the interaction picture. In quantum optics this is referred to as a rotating frame. In the lab frame, we also have sinusoidal time dependence $a \sim a e^{i \omega t}$ and so forth, where $\omega$ is the frequency of the relevant mode, which causes the squeezed quadrature to oscillate between $X$ and $Y$ (or $h$ and $\pi_h$). In the right panel of Fig. \ref{fig:time-dep}, we show how this can lead to an oscillating variance in the detector output. The variance oscillates between sub-vacuum and super-vacuum values. The conclusion is the same: the sub-vacuum parts require a quantum signal mode to explain, but the super-vacuum parts can be explained by a purely classical signal.

\section{LIGO and squeezed gravitons}
\label{ligo}

Finally, let's explicitly connect our simple models above to the suggestion in \cite{Guerreiro:2019vbq,Parikh:2020fhy} that a highly squeezed gravitational signal could produce a large amount of ``quantum noise''. It is true that a squeezed state produces a lot of extra noise (i.e., the $\Delta \pi_h > 1/2$ above, which oscillates into the detector signal as in Fig. \ref{fig:time-dep}). However, the majority of this can be explained with a classical random signal, as we just discussed. It is only the sub-vacuum part that requires a quantum explanation. But this part is not observable in a real detector. We now make this precise by working out some numbers for LIGO. Despite the non-observability, the calculation is interesting, because it shows how the sub-vacuum graviton noise can be in principle transduced into a macroscopic detector.

In continuous linear detectors, the noise is characterized by the noise power spectral density (PSD),
\be
S_{hh}(\nu) = \int dt\ e^{-i \nu t} \braket{ h(t) h(0)}.
\ee
The actual observed noise power has a component both from the signal (i.e., the gravitational field) as well as noise due to the detector itself, in particular noise from the quantum readout as in both the photodiode and amplitude detectors discussed above. We can express this in terms of the optical mode's variables $X = (a + a^\dag)/\sqrt{2}$ and $Y = -i (a - a^\dag)/\sqrt{2}$. In the usual discussion of LIGO-like detectors \cite{Beckey:2023shi}, one works in a coordinate frame where a suspended mirror with center-of-mass variables $x,p$ couples to this optical mode through an optomechanical coupling $V = G x X$. A gravitational wave in this frame acts as an effective force $F^{\rm sig} \sim m L \ddot{h}$, where $L$ is the baseline of the interferometer. Gravitational waves are detected by the signal driving the mirror motion, which in turn drives the optical field, which is then measured, say in the $Y$ basis. The noise power spectral density of this readout is given by an expression like
\be
\label{SYY}
S^{\rm out}_{YY} = |\chi_{YY}|^2 S_{YY}^{\rm in} + |\chi_{YX}|^2 S_{XX}^{\rm in} + \cdots,
\ee
where the right-hand side contains a bunch of ``input'' noise PSDs $S_{ij}$ and response functions $\chi_{ij}$. These come from the input laser fluctuations, thermal forces on the mirror, and also include the gravitational noise of interest here. The question is: can a squeezed (or otherwise quantum) gravitational signal \emph{reduce} the detector output $S_{YY}^{\rm out,quantum}$ below the minimum possible noise $S_{YY}^{\rm out,classical}$ which is consistent with a classical random gravitational signal?

To analyze this, we need to compute the noise PSD $S_{YY}^{\rm out}$. This can be done using a standard method called the input-output formalism \cite{Beckey:2023shi,clerk2010introduction}, which essentially amounts to solving the linear equations of motion of the detector and assuming the level of noise injected by the various noise sources. In the input-output formalism, the phase and amplitude of the light coming out of the cavity is related to that coming in through the input-output relations
\be
X_{\rm out} = X_{\rm in} + \sqrt{\kappa} X, \ \ \ Y_{\rm out} = Y_{\rm in} + \sqrt{\kappa} Y.
\ee
To use these, we need to solve the equations of motion for the cavity light, i.e., for $X$ and $Y$. The Heisenberg equations for the mirror motion take the form
\begin{align}
\begin{split}
\label{EOM-mech}
\dot{x} & = m \omega^2 x \\
\dot{p} & = - \gamma_{m} p - \gamma_{h} p + F^{\rm in} + F^{\rm sig} + G X,
\end{split}
\end{align}
The last term gives the coupling of the mirror motion to the optical readout. The mirror motion is harmonic with mass $m \approx 40~{\rm kg}$ and frequency $\omega \approx 1~{\rm Hz}$. The coefficients $\gamma_m$ and $\gamma_h$ represent damping of the mirror respectively from coupling to any ambient thermalizing bath like gas (which exerts the random thermal force $F^{\rm in}$), and damping of the mirror from coupling to the gravitational field (through $F^{\rm sig}$). Both of these terms are usually neglected in LIGO; the former because it is subdominant to the noise coming from the coupling to $X$, and the latter because it is astronomically small. However, loss of information via $\gamma_h$ will play a key role in what follows: it is the analogue to the unmonitored port in the beamsplitter model of Sec. 4. 

In addition to the equations of motion for the mirror, we have the equations of motion for the optical field in the cavity,
\begin{align}
\begin{split}
\dot{X} & = -\frac{\kappa}{2} X + \sqrt{\kappa} X_{\rm in} \\
\dot{Y} & = -\frac{\kappa}{2} Y + \sqrt{\kappa} Y_{\rm in} + G x.
\end{split}
\end{align}
Here $\kappa$ is the rate at which photons leak into and out of the cavity, i.e., the coupling to the optical modes outside of the cavity, in particular the laser field. We are in the frame co-rotating with the cavity and assume zero detuning between the laser and cavity mode. 

The equations of motion can be solved easily by moving to the frequency domain with $Y(t) = \int d\nu e^{i \nu t} Y(\nu)$ and so forth. Inserting the solutions into the input-output relations, we obtain the following expression for the phase of the light coming out:
\be
\label{Yout}
Y_{\rm out} = \chi_{YY} Y_{\rm in} + \chi_{YX} X_{\rm in} + \chi_{YF} F_{\rm in} + \chi_{Yh} h_{\rm in},
\ee
where the response functions are
\begin{equation}
\begin{aligned}
\label{chis}
\chi_{YY} & = \frac{-i \nu - \kappa/2}{-i \nu + \kappa/2} = e^{i \phi}, & \chi_{YX}  & = \kappa G^2 \chi_c^2 \chi_m \\
\chi_{YF} & = \sqrt{\kappa} G \chi_c \chi_m,  & \chi_{Yh} & = \sqrt{\kappa} G \chi_c \chi_m m^2 L^2 \nu^4,
\end{aligned}
\end{equation}
in terms of the basic cavity and mechanical response functions
\be
\chi_c(\nu) = \frac{1}{-i \nu + \kappa/2}, \ \ \ \chi_m(\nu) = \frac{1}{m [(\omega^2 - \nu^2) - i \gamma \nu]}.
\ee
The total mechanical damping is the sum of the usual thermal damping term as well as the damping from loss into the gravitational field itself
\be
\gamma = \gamma_m + \gamma_h.
\ee
In practice, in LIGO, the thermal noise effects (represented by $F_{\rm in}$ and $\gamma_m$) are subdominant to the quantum noise (coming from the $Y_{\rm in}$ and $X_{\rm in}$ terms). All of these are of course hugely dominant over the gravitational noise in usual signals, but here we are interested in an extreme case where the gravitational noise is important.

Finally, we can use these solutions to compute the detector's noise PSD. This can be easily obtained from Eq. \eqref{Yout} by what is called the Weiner-Khinchin theorem, which essentially says that $S_{xx}(\nu) \propto \braket{ |x(\nu)|^2 }$ for any observable $x$ (see \cite{Beckey:2023shi} for a detailed treatment). Here this gives Eq. \eqref{SYY}, i.e.,
\be
S^{\rm out}_{YY} = |\chi_{YY}|^2 S_{YY}^{\rm in} + |\chi_{YX}|^2 S_{XX}^{\rm in} + |\chi_{YF}|^2 S_{FF}^{\rm in} + |\chi_{Yh}|^2 S^{\rm in}_{hh}.
\ee
The input noise from the laser can be modeled as simple vacuum fluctuations, where
\be
\label{vacin}
S_{YY}^{\rm in} = S_{XX}^{\rm in} = 1/2.
\ee
This $1/2$ is exactly the vacuum-state variance of the harmonic oscillator variables used in the simple examples in the main text. In LIGO currently the situation is slightly more sophisticated since squeezed vacuum is inserted \cite{Jia:2024epj}, but this is not going to change anything we are about to say, so for simplicity we will stick with standard vacuum input. The thermal noise is modeled as white noise
\be
S_{FF}^{\rm in} = 2( n_T(\omega) + 1) m  \omega \gamma_m,
\ee
where $n_T(\omega) = 1/(1-e^{\omega/T})$ is the Bose-Einstein factor at the mechanical frequency and the $+1$ term represents vacuum fluctuations of the bath.

Now we want to compare the situation in which the gravitational field is purely classical or quantized. Following the main text, the idea is that for the quantized case, the detector output noise $S_{YY}^{\rm out}$ can be decreased below what is possible in the classical case, for example by transducing sub-vacuum squeezed noise from $h$ into $Y$. There are two main differences between the quantum and classical radiation field treatments. One is that in the classical case, the input noise $S_{hh}^{\rm in} = 0$ is possible, whereas in the quantized case there is always vacuum noise. The other is that in the classical case, there can be zero damping $\gamma_h = 0$, for the same reason: unless $h$ is quantized, there is no amplitude for the mirror to spontaneously dump energy into the field. These effects are related, as encoded in the fluctuation-dissipation theorem evaluated at zero temperature, which here reads \cite{clerk2010introduction,whittle2023unification}
\be
\label{FD}
\gamma_h = x_0^2 m^2 L^2 \omega^4 \left[ S^{\rm in}_{hh}(\omega) - S^{\rm in}_{hh}(-\omega) \right],
\ee
where $x_0 = 1/\sqrt{2 m \omega}$ is the zero-point uncertainty of the mirror position. For classical $h$, the right-hand side is strictly equal to zero (the PSD is symmetric), while in the quantized case it equals $1$, from vacuum fluctuations (i.e., from the term where you commute $[b,b^\dag]=1$). Using these two observations, we can compute the ratio of the detector's output noise in the classical and quantum cases:
\be
\label{ratio-app}
\frac{S_{YY}^{\rm out, quantum}}{S_{YY}^{\rm out, classical}} = \frac{1 + |\chi_{YX}|^2 + 2 |\chi_{YF}|^2 S_{FF}^{\rm in} + 2 |\chi_{Yh}|^2 e^{-2r} S^{\rm vac}_{hh}}{1  + |\chi_{YX}|_{\gamma_h = 0}^2 + 2 |\chi_{YF}|_{\gamma_h=0}^2 S_{FF}^{\rm in} }.
\ee
Here we used the vacuum laser values from Eq. \eqref{vacin}, assumed a general squeezed graviton state with squeezing $r$, and we are specializing to the case of noiseless classical fluctuations in the denominator. The physics is directly analogous to our discussion of a quantum or classical local oscillator beamsplitter around Eqs. \eqref{Vint} and \eqref{Vcl}, where here the mechanical element interacting with the gravitational wave is playing the role of the beamsplitter.

We can get a very transparent approximate answer out of Eq. \eqref{ratio-app}. Let's focus on LIGO, which observes at frequencies around $\nu_* \approx 1~{\rm kHz}$. This scale is much larger than the mechanical resonance and cavity loss rates $\nu_* \gg \omega, \kappa$ which considerably simplifies the response functions in Eq. \eqref{chis}. Also, the photon-mirror coupling $G \propto \sqrt{P_L}$ depends on the input laser power and is therefore tuneable; LIGO chooses to tune it to $G^2 = [2 \kappa |\chi_c(\nu_*)|^2 |\chi_m(\nu^*)|]^{-1}$. This has the effect of minimizing the total quantum noise at $\nu_*$ (achieving the ``Standard Quantum Limit'' at $\nu_*$), and sets $|\chi_{YX}|^2 = 1$.\footnote{For the connoisseurs: doing this relies on knowing the value of $\gamma$. Here we are assuming that we know the full $\gamma = \gamma_m + \gamma_h$, for example through some calibration; if we instead used $\gamma = \gamma_m$ this would give a further $O(M_{\rm pl}^{-2})$ correction.} Put together this simplifies the output noise to
\be
S_{YY}^{\rm out} = 1 + |\chi_m| S_{FF}^{\rm in} + |\chi_m| m^2 L^2 \nu^4 S_{hh}^{\rm in},
\ee
in either the classical or quantum case, where here we are focusing on the noise PSD evaluated at the signal band $\nu_* \approx 1~{\rm kHz}$. Now the key is that $|\chi_m|$ depends on the damping parameter $\gamma_h$, since it involves the response of the mirror to external forces [see Eq. \eqref{chis}]. A simple Taylor expansion shows that it depends linearly on $\gamma_h \ll \gamma_m$,
\be
|\chi_m| = |\chi_m|_0 \left( 1 - \gamma_h \gamma_m m^2 \nu^2 |\chi_m|^2_0 + \cdots \right).
\ee
Using this and the scaling from the fluctuation-dissipation theorem [Eq. \eqref{FD}], we can expand the ratio in Eq. \eqref{ratio-app} to lowest order in $1/M_{\rm pl}^2$. This gives
\be
\frac{S_{YY}^{\rm out, quantum}}{S_{YY}^{\rm out, classical}} \approx 1 - \gamma_h \gamma_m m^2 \nu^2 |\chi_m|_0^3 S_{FF}^{\rm in} + |\chi_m|_0 m^2 L^2 \nu^4 e^{-2 r} S_{hh}^{\rm vac}.
\ee
The basic structure of this answer is exactly the same as the simple beamsplitter case discussed in the main text in Eq. \eqref{deltaXpert}. The final term is an additional noise coming from the input graviton state. However, the second term shows a \emph{loss} of noise at the same order; this is the low-noise graviton state being transduced into the mechanical state. Just like the beamsplitter example, however, the positive contribution can be made arbitrarily small with sufficient squeezing $r \to \infty$, while the reductive term is independent of $r$. Thus, a sufficiently squeezed graviton state can actually act to reduce the observed detector output below what is possible with a purely classical gravitational wave.

Unfortunately, it would be impossible to detect this signature of quantum gravity. To see it would require that the two correction terms here add up to some total which is negative and order one (which physically means that they subtract a meaningful fraction of the intrinsic detector quantum noise, represented by the 1 term). In the extreme case of a completely squeezed graviton state $r \to \infty$ we can entirely drop the positive term, and estimate the size of the noise reduction. For LIGO, this works out to be of order
\be
\frac{S_{YY}^{\rm out, quantum}}{S_{YY}^{\rm out,classical}} \approx  1 - \gamma_h \gamma_m m^2 \nu^2 |\chi_m|^3_0 S_{FF}^{\rm in} + \cdots \approx 1 - 10^{-43},
\ee
where we used $\nu = 1~{\rm kHz}$, $m = 40~{\rm kg}$, $L = 4~{\rm km}$, $\gamma_m = 10^{-6}~{\rm Hz}$ (a very conservative estimate; $10^{-8}$ is probably more accurate), the vacuum graviton PSD $S_{hh}^{\rm vac}(\nu) \approx \nu/M_{\rm pl}^2$ in the fluctuation-dissipation calculation of $\gamma_h$ [see Eq. \eqref{FD}], and set the mechanical bath to $T = 300~{\rm K}$. The conclusion is clear: there is no chance of seeing this quantum-only signature in a real detector, even in the extreme case studied here where the gravitational wave is infinitely squeezed.

\section{Comments}

Given that all the fields of nature other than gravity are known to be quantized, perhaps the most natural guess is that gravity will be quantized similarly. However, given the difficulties encountered when trying to extrapolate this quantization to very high energies, it is worth seeking definitive experimental evidence that the field is quantized in the first place. A natural place to start looking is in the gravitational radiation field, now detectable by a variety of devices. Unfortunately, the considerations above suggest that any measurement of gravitational radiation with a realistic detector will be unable to distinguish classical gravitational waves from quantized ones. The remaining options appear to be tabletop experiments \cite{Carney:2018ofe}, which probe the quantization of the Newton potential (i.e., virtual graviton exchange), or measurements of the CMB \cite{Green:2020whw}.

\section*{Acknowledgements}

The scientific content of this paper is largely based on joint work with Valerie Domcke and Nick Rodd \cite{Carney:2023nzz}, and I am grateful to them for the wonderful collaboration we have had on this topic. This writeup is a slightly expanded version of a talk prepared for Gordon Semenoff's 70th birthday festschrift held at the Universit\'e de Montreal, July 2023. Gordon and I have been thinking about quantization effects in low-energy gravity since my very first paper on the topic \cite{Carney:2017jut}. I would like to congratulate him on his illustrious career, and thank him heartily for both our scientific collaboration and perhaps more importantly his mentorship over the years since we overlapped at UBC. My work at Berkeley Lab is supported by the U.S. DOE, Office of High Energy Physics, under Contract No. DEAC02-05CH11231, by DOE Quantum Information Science Enabled Discovery (QuantISED) for High Energy Physics grant KA2401032, and by the Heising-Simons Foundation.

\bibliographystyle{utphys-dan}
\bibliography{comments-on-graviton-detection-gordon-fest.bib}

\providecommand{\href}[2]{#2}\begingroup\raggedright\begin{thebibliography}{10}

\bibitem{Carney:2023nzz}
D.~Carney, V.~Domcke, and N.~L. Rodd, ``{Graviton detection and the
  quantization of gravity},''
  \href{http://dx.doi.org/10.1103/PhysRevD.109.044009}{{\em Phys. Rev. D}
  {\bfseries 109} no.~4, (2024) 044009},
  \href{http://arxiv.org/abs/2308.12988}{{\ttfamily arXiv:2308.12988
  [hep-th]}}.

\bibitem{Donoghue:2022eay}
J.~F. Donoghue, {\em {Quantum General Relativity and Effective Field Theory}}.
\newblock 2023.
\newblock \href{http://arxiv.org/abs/2211.09902}{{\ttfamily arXiv:2211.09902
  [hep-th]}}.

\bibitem{Boughn:2006st}
S.~Boughn and T.~Rothman, ``{Aspects of graviton detection: Graviton emission
  and absorption by atomic hydrogen},''
  \href{http://dx.doi.org/10.1088/0264-9381/23/20/006}{{\em Class. Quant.
  Grav.} {\bfseries 23} (2006) 5839--5852},
  \href{http://arxiv.org/abs/gr-qc/0605052}{{\ttfamily arXiv:gr-qc/0605052}}.

\bibitem{Dyson:2013hbl}
F.~Dyson, ``{Is a graviton detectable?},''
  \href{http://dx.doi.org/10.1142/S0217751X1330041X}{{\em Int. J. Mod. Phys. A}
  {\bfseries 28} (2013) 1330041}.

\bibitem{Palessandro:2024ria}
A.~Palessandro, ``{Graviton-Photon Oscillations as a Probe of Quantum
  Gravity},'' \href{http://arxiv.org/abs/2405.01407}{{\ttfamily
  arXiv:2405.01407 [gr-qc]}}.

\bibitem{Tobar:2023ksi}
G.~Tobar, S.~K. Manikandan, T.~Beitel, and I.~Pikovski, ``{Detecting single
  gravitons with quantum sensing},''
  \href{http://arxiv.org/abs/2308.15440}{{\ttfamily arXiv:2308.15440
  [quant-ph]}}.

\bibitem{Shenderov:2024rup}
V.~Shenderov, M.~Suppiah, T.~Beitel, G.~Tobar, S.~K. Manikandan, and
  I.~Pikovski, ``{Stimulated absorption of single gravitons: First light on
  quantum gravity},'' \href{http://arxiv.org/abs/2407.11929}{{\ttfamily
  arXiv:2407.11929 [gr-qc]}}.

\bibitem{Kafri:2014zsa}
D.~Kafri, J.~M. Taylor, and G.~J. Milburn, ``{A classical channel model for
  gravitational decoherence},''
  \href{http://dx.doi.org/10.1088/1367-2630/16/6/065020}{{\em New J. Phys.}
  {\bfseries 16} (2014) 065020},
  \href{http://arxiv.org/abs/1401.0946}{{\ttfamily arXiv:1401.0946
  [quant-ph]}}.

\bibitem{Tilloy:2018tjp}
A.~Tilloy, ``{Binding quantum matter and space-time, without romanticism},''
  \href{http://dx.doi.org/10.1007/s10701-018-0224-6}{{\em Found. Phys.}
  {\bfseries 48} no.~12, (2018) 1753--1769},
  \href{http://arxiv.org/abs/1802.03291}{{\ttfamily arXiv:1802.03291
  [physics.hist-ph]}}.

\bibitem{Oppenheim:2023izn}
J.~Oppenheim and Z.~Weller-Davies, ``{Covariant path integrals for quantum
  fields back-reacting on classical space-time},''
  \href{http://arxiv.org/abs/2302.07283}{{\ttfamily arXiv:2302.07283 [gr-qc]}}.

\bibitem{Glauber:1963tx}
R.~J. Glauber, ``{Coherent and incoherent states of the radiation field},''
  \href{http://dx.doi.org/10.1103/PhysRev.131.2766}{{\em Phys. Rev.} {\bfseries
  131} (1963) 2766--2788}.

\bibitem{sudarshan1963equivalence}
E.~Sudarshan, ``Equivalence of semiclassical and quantum mechanical
  descriptions of statistical light beams,'' {\em Physical Review Letters}
  {\bfseries 10} no.~7, (1963) 277.

\bibitem{mandel1995optical}
L.~Mandel and E.~Wolf, {\em Optical Coherence and Quantum Optics}.
\newblock Cambridge University Press, 1995.

\bibitem{Guerreiro:2019vbq}
T.~Guerreiro, ``{Quantum Effects in Gravity Waves},''
  \href{http://dx.doi.org/10.1088/1361-6382/ab9d5d}{{\em Class. Quant. Grav.}
  {\bfseries 37} no.~15, (2020) 155001},
  \href{http://arxiv.org/abs/1911.11593}{{\ttfamily arXiv:1911.11593
  [quant-ph]}}.

\bibitem{Parikh:2020fhy}
M.~Parikh, F.~Wilczek, and G.~Zahariade, ``{Signatures of the quantization of
  gravity at gravitational wave detectors},''
  \href{http://dx.doi.org/10.1103/PhysRevD.104.046021}{{\em Phys. Rev. D}
  {\bfseries 104} no.~4, (2021) 046021},
  \href{http://arxiv.org/abs/2010.08208}{{\ttfamily arXiv:2010.08208
  [hep-th]}}.

\bibitem{https://doi.org/10.1002/andp.19053220607}
A.~Einstein, ``Über einen die erzeugung und verwandlung des lichtes
  betreffenden heuristischen gesichtspunkt,''
  \href{http://dx.doi.org/https://doi.org/10.1002/andp.19053220607}{{\em
  Annalen der Physik} {\bfseries 322} no.~6, (1905) 132--148}.

\bibitem{mehta1967diagonal}
C.~Mehta, ``Diagonal coherent-state representation of quantum operators,'' {\em
  Physical Review Letters} {\bfseries 18} no.~18, (1967) 752.

\bibitem{loudon1987squeezed}
R.~Loudon and P.~L. Knight, ``Squeezed light,'' {\em Journal of modern optics}
  {\bfseries 34} no.~6-7, (1987) 709--759.

\bibitem{short1983observation}
R.~Short and L.~Mandel, ``Observation of sub-poissonian photon statistics,''
  \href{http://dx.doi.org/10.1103/PhysRevLett.51.384}{{\em Phys. Rev. Lett.}
  {\bfseries 51} (Aug, 1983) 384--387}.

\bibitem{Jia:2024epj}
W.~Jia {\em et~al.}, ``{LIGO operates with quantum noise below the Standard
  Quantum Limit},'' \href{http://arxiv.org/abs/2404.14569}{{\ttfamily
  arXiv:2404.14569 [gr-qc]}}.

\bibitem{Beckey:2023shi}
J.~Beckey, D.~Carney, and G.~Marocco, ``{Quantum measurements in fundamental
  physics: a user's manual},''
  \href{http://arxiv.org/abs/2311.07270}{{\ttfamily arXiv:2311.07270
  [hep-ph]}}.

\bibitem{clerk2010introduction}
A.~A. Clerk, M.~H. Devoret, S.~M. Girvin, F.~Marquardt, and R.~J. Schoelkopf,
  ``Introduction to quantum noise, measurement, and amplification,'' {\em
  Reviews of Modern Physics} {\bfseries 82} no.~2, (2010) 1155--1208.

\bibitem{whittle2023unification}
C.~Whittle, L.~McCuller, V.~Sudhir, and M.~Evans, ``Unification of thermal and
  quantum noises in gravitational-wave detectors,'' {\em Physical Review
  Letters} {\bfseries 130} no.~24, (2023) 241401.

\bibitem{Carney:2018ofe}
D.~Carney, P.~C.~E. Stamp, and J.~M. Taylor, ``{Tabletop experiments for
  quantum gravity: a user\textquoteright{}s manual},''
  \href{http://dx.doi.org/10.1088/1361-6382/aaf9ca}{{\em Class. Quant. Grav.}
  {\bfseries 36} no.~3, (2019) 034001},
  \href{http://arxiv.org/abs/1807.11494}{{\ttfamily arXiv:1807.11494
  [quant-ph]}}.

\bibitem{Green:2020whw}
D.~Green and R.~A. Porto, ``{Signals of a Quantum Universe},''
  \href{http://dx.doi.org/10.1103/PhysRevLett.124.251302}{{\em Phys. Rev.
  Lett.} {\bfseries 124} no.~25, (2020) 251302},
  \href{http://arxiv.org/abs/2001.09149}{{\ttfamily arXiv:2001.09149
  [hep-th]}}.

\bibitem{Carney:2017jut}
D.~Carney, L.~Chaurette, D.~Neuenfeld, and G.~W. Semenoff, ``{Infrared quantum
  information},'' \href{http://dx.doi.org/10.1103/PhysRevLett.119.180502}{{\em
  Phys. Rev. Lett.} {\bfseries 119} no.~18, (2017) 180502},
  \href{http://arxiv.org/abs/1706.03782}{{\ttfamily arXiv:1706.03782
  [hep-th]}}.

\end{thebibliography}\endgroup

\end{document}